\documentclass[11pt,preprint]{aastex}
\begin{document}

\newcommand{\kms}{$\,\mbox{km}\,\mbox{s}^{-1}$}
\newcommand{\etal}{et al.}
\newcommand{\LCDM}{$\Lambda$CDM}
\newcommand{\Om}{\ensuremath{\Omega_m}}
\newcommand{\Ob}{\ensuremath{\Omega_b}}
\newcommand{\ob}{\ensuremath{\omega_b}}
\newcommand{\wb}{\ensuremath{\omega_b}}
\newcommand{\On}{\ensuremath{\Omega_{\nu}}}
\newcommand{\fnu}{\ensuremath{f_{\nu}}}
\newcommand{\mnu}{\ensuremath{m_{\nu}}}
\newcommand{\OL}{\ensuremath{\Omega_{\Lambda}}}
\newcommand{\LL}{\ensuremath{\ell}}
\newcommand{\apk}{\ensuremath{A_{1:2}}}
\newcommand{\rpk}{\ensuremath{R_{1:2}}}
\newcommand{\HI}{H{\sc i}\ }


\title{Confrontation of MOND Predictions \\ 
with WMAP First Year Data}

\author{Stacy S.~McGaugh} 
\affil{Department of Astronomy, University of Maryland}
\affil{College Park, MD 20742-2421}    
\email{ssm@astro.umd.edu}

\begin{abstract}
I present a model devoid of non-baryonic cold dark matter (CDM) 
which provides an acceptable fit to the WMAP data for the
power spectrum of temperature fluctuations in the cosmic 
background radiation (CBR).  An {\it a priori\/} prediction of
such no-CDM models was a
first-to-second peak amplitude ratio $\apk \approx 2.4$.
WMAP measures $\apk = 2.34 \pm 0.09$.
The baryon content is the dominant factor in fixing this ratio;
no-CDM models which are consistent with the WMAP data
are also consistent with constraints on the baryon density from 
the primordial abundances of $^2$H, $^4$He, and $^7$Li.
However, in order to match the modest width of the acoustic 
peaks observed by WMAP, a substantial neutrino mass is implied:
$m_{\nu} \approx 1$~eV.  Even with such a heavy neutrino,
structure is expected to form rapidly under the influence of 
MOND.  Consequently, the epoch of reionization should occur 
earlier than is nominally expected in \LCDM.  This prediction 
is realized in the polarization signal measured by WMAP.  
An outstanding test is in the amplitude of
the third acoustic peak.  Experiments which probe high-\LL\ 
appear to favor a third peak which is larger than predicted
by the no-CDM model.
\end{abstract}


\section{Introduction}

The past decade has seen a remarkable convergence towards a
standard cosmological model, \LCDM.  A great variety of different
and independent data seem to point to very nearly the same set of 
cosmological parameters.  The high precision 
measurement of the temperature fluctuations in the microwave
background by the WMAP satellite (Bennett \etal\ 2003)
has given further impetus to this picture.

While \LCDM\ does provide a fairly consistent cosmology,
there do remain nagging problems, some of
which are potentially quite serious.  Perhaps the most 
embarrassing\footnote{For those of us who found the Inflationary
solution to the coincidence problem so satisfying, the presently
inferred acceleration of the expansion of the universe --- which makes
the coincidence problem worse than it had been in an open universe ---
is also rather embarrassing.}
aspect of our modern cosmology is the dominance of invisible
components.  Dark matter and dark energy comprise $\gtrsim 95\%$ 
of the mass-energy content of the universe, yet we have only ideas
about what they are.

In the context of cosmology, it is essential for the dark matter to be
non-baryonic (in order to satisfy $\Om \gg \Ob$) and dynamically cold
(in order to grow structure).  This led to the development of the
Cold Dark Matter paradigm, in conjunction with the suggestion from
particle physics that a class of supersymmetric, weakly interacting
massive particles (WIMPs) might have the right properties to be the
CDM.  Our modern cosmology absolutely requires CDM,
but as yet we have no compelling laboratory evidence that such 
particles actually exist.

The CDM hypothesis faces serious problems on the scale of individual
galaxies (Sellwood \& Kosowsky 2001).  One widely debated problem is
that dark matter halos should have steep central density cusps 
(e.g., Navarro, Frenk, \& White 1997) but appear not to
(e.g., de Blok, Bosma, \& McGaugh 2003; Swaters \etal\ 2003;
for further references see McGaugh, Barker, \& de Blok 2003).
While this is an important point, the cusp problem itself is only 
one aspect of the severe fine-tuning problems one encounters
in trying to understand the systematic properties of the rotation
curves of spiral galaxies (McGaugh \& de Blok 1998a; McGaugh 2004).

The difficulty for CDM encountered in the dynamics of individual
galaxies stems from the close coupling observed between baryonic
and dark matter components.  The distribution of the observed
baryons is completely predictive of that of the dynamically
dominant dark matter (Sancisi 2003, McGaugh 1999a; 2000a; 2004).
One does not naturally expect, and finds difficult to impart
(McGaugh \& de Blok 1998a), such a close coupling between 
a small, dynamically cold, thin disk of baryons and a large,
dynamically hot, quasi-spherical halo of non-baryonic dark matter.

Into this troubling mix comes the surprising success of the Modified 
Newtonian Dynamics (MOND) proposed by Milgrom (1983).
Long known to provide good fits to the rotation curves of high surface
brightness spiral galaxies (e.g., Begeman, Broeils, \& Sanders 1991;
Sanders 1996), MOND also works in systems ranging from 
tiny dwarf Spheroidals\footnote{Ursa Minor and Draco are often cited
as problematic cases for MOND.  This is true, but it also happens that
for both of these cases, it is unclear (at $< 1 \sigma$) which MOND
mass estimator is appropriate:  both are very close to the dividing line
between internal and external field domination.  For the other eight cases
with adequate data (Carina, Fornax, LGS3, Leo I, Leo II, Sagitarius, 
Sculptor, and Sextans) where the choice of mass estimator is not 
ambiguous, the results are consistent with MOND.}
(Sanders \& McGaugh 2002) to giant Ellipticals (Milgrom \& Sanders 2003).
While there is a genuine puzzle in rich clusters of galaxies 
(Aguirre, Schaye, \& Quataert 2001), it is not obvious that this 
problem is more serious than those faced by CDM.
One should also bear in mind the surprising success of the very specific
{\it a priori\/} predictions made for low surface brightness (LSB) galaxies by
Milgrom (1983), which were realized in great detail (McGaugh \& de Blok 1998b).
All of the things which are so puzzling about the dynamics of these systems
from a conventional point of view stem
fundamentally from their adherence to the MOND force law.
This should not happen if CDM is correct.

As stressed by McGaugh \& de Blok (1998b), the dynamical data admit
only two interpretations.  Either MOND is correct (presumably as the
appropriate limit of some more general theory), or the processes of
galaxy formation somehow lead to MOND-like behavior.  The latter
case is discussed elsewhere (McGaugh 1999a; 2000a; 2004), and provides
quite stringent empirical constraints on conventional galaxy formation theory.
However, should MOND be correct, how else might we tell?
The CBR provides a further opportunity to distinguish between the very
different cases of a ``preposterous'' universe
filled with dark matter and dark energy and an even more mysterious one in
which the mass discrepancies observed in extragalactic systems are
caused by a modification of dynamical laws.  

MOND itself is not a generally covariant theory.
This makes it impossible (not merely daunting!) to derive a proper 
cosmology and compute the power spectrum of the CBR.
However, as is often the case in such situations, 
a simple {\it ansatz\/} can be very illuminating.

McGaugh (1999b) made the {\it ansatz\/}  that the acceleration scale
$a_0$ of MOND does not vary with time.  This is a common assumption
(Felten 1984; Sanders 1998) with the consequence that the early universe
is not in the MOND regime.  In this case, things are normal until sufficiently
late times that the usual early universe results (e.g., Big Bang
Nucleosynthesis, BBN) are retained.  

The point of MOND is to obviate the need for dark matter.  
If it is essentially correct, then CDM does not exist.
This suggests a minimum difference approach: 
if MOND is active in the early universe,
then presumably the effects on the CBR would be larger than what we
would infer under the {\it ansatz}.  If MOND is not active
at the time of recombination, and there
is no CDM, then the CBR power spectrum should have a distinctive
shape.  Without CDM, baryonic drag dominates, and one should
see only the resultant damping, with each peak being lower in amplitude
than the preceding one.  If CDM does exist in the required mass density,
it must provide some net forcing term for the acoustic oscillations
which counteracts the baryonic drag:  one should never see a pure 
damping spectrum in a \LCDM\ universe.

Since we use only conventional physics under our {\it ansatz}, MOND
itself is not tested.  Thus we can not exclude MOND on the basis of any
CBR data simply for lack of a definite prediction.  However, we can 
potentially exclude the existence of CDM.  It is very important to have
such a test, for once we have convinced ourselves of the need for an
all-pervasive, invisible, and perhaps undetectable mass component,
it is virtually impossible to falsify the notion should it happen to be wrong
(Davis \etal\ 1992).

To this end, McGaugh (1999b) proposed the no-CDM model as a stand-in
for MOND.  This provides a minimal first approximation of what we
might expect in MOND.  Of course, it should fail at some level.  
We have known for
a long time that a purely Newtonian, baryonic model fails --- we must have
either CDM or MOND.  Even under the {\it ansatz\/} that MOND is not
important at the time of recombination, there can be late time effects
which modify the observed CBR.  Given these considerations,
McGaugh (1999b) also discussed some of the deviations from
a pure no-CDM model which are likely to be caused by MOND.
For example, rapid structure formation is natural to
MOND (Sanders 1998; McGaugh 1999c).  This
should lead to early reionization and consequently a large polarization 
signal.  It may also enhance the Integrated Sachs-Wolfe (ISW) effect.
These are indicators of MOND-like physics
beyond the simple no-CDM model. 

The amplitudes of the peaks expected in the CBR power spectrum
in the absence of CDM is discussed in \S 2.  These are tied to the baryon
density, for which independent constraints from BBN are critical.  A 
specific no-CDM model which fits the WMAP data is given, and its predictions
for the third and subsequent peaks are checked against higher \LL\ data from
other experiments.  Features of the CBR power spectrum which might
suggest MOND-like physics beyond the simple no-CDM model are discussed
in \S 3.  Further predictions which could help to distinguish between CDM
and MOND are discussed in \S 4.

\section{The No-CDM Model}

\subsection{The Amplitude of the Second Peak}

The difference between a standard \LCDM\ model and the MOND 
inspired no-CDM model turns out to be rather subtle.
The first difference is a modest one in the amplitude of the second peak
relative to the first: the second peak should be somewhat smaller
without CDM for the same baryon density.
Quite high quality data are required to distinguish between CDM and MOND
when only the first two peaks are observed.

McGaugh (1999b) suggested two robust measures.  The first is the 
ratio of the absolute amplitude of the first to second peak:
\begin{equation}
\apk = \frac{A_1}{A_2} = \frac{C_{{\LL},1}}{C_{{\LL},2}},
\end{equation}
where $C_{{\LL},N}$ is the peak amplitude of the $N^{th}$ peak.
(A similar but inverted notation was adopted by Hu \etal\ 2001:
their $H_2 = \apk^{-1}$.)
The second measure is the peak amplitude ratio relative to the
intervening trough ($C_{{\LL},t}$):
\begin{equation}
\rpk = \frac{R_1}{R_2} =
 \frac{C_{{\LL},1}-C_{{\LL},t}}{C_{{\LL},2}-C_{{\LL},t}}.
\end{equation}
Though both of the ratios \apk\ and \rpk\ contain much the same 
information, the peak amplitude ratio relative to the intervening trough
(\rpk) gives a stronger segregation between the model predictions.
However, it is sensitive to uncertainties in the trough amplitude,
which is harder to measure than the peaks.  

The definition of these ratios is illustrated in Fig.~1 with the no-CDM models
of McGaugh (1999b) and the WMAP peak location estimates of
Page \etal\ (2003).  Note that the no-CDM models are consistent with
the WMAP peak amplitude estimates, having very similar
peak ratios for plausible baryon densities.  Coeval
\LCDM\ models generically had larger second peaks.

In order to clarify the \LCDM\ expectation value for \apk\ and \rpk\ as it
existed prior to CBR constraints, I have produced new models using 
``\LCDM\ 1999'' parameters (e.g., Turner 1999). 
These are completely ``vanilla'' models
with reasonable parameters for the time, i.e., $\Om = 0.3$, $\OL = 0.7$,
and $n = 1$.  The only significant difference from the \LCDM\ models 
discussed by McGaugh (1999b), or for that matter, from current \LCDM\ 
models, is the baryon density.  This I fix to the value given by Tytler \etal\ 
(2000): $\ob \equiv \Ob h^2 = 0.019$.  
This is a little higher than the baryon densities
I had considered in 1999, which were chosen to sample the range suggested
by prior BBN reviews (Walker \etal\ 1991; Copi, Schramm, \& Turner 1995).
A higher baryon density produces a lower second peak which is more 
favorable for \LCDM\ in the context of subsequent
CBR observations.  This being the case, I also give a model with
$\ob = 0.0214$, which is the 95\% c.l.\ upper limit of Tytler \etal\ (2000).
The result is an upper limit on the \LCDM\ prior expectation for the
amplitude ratio of $\apk < 2.06$ (Table~1).  This is as generous as it is
reasonable to be to the \LCDM\ paradigm as it existed before stringent CBR
constraints, when there was no cause to consider higher baryon densities
or exotic effects like a running tilt to the power spectrum.

The first data constraining
the first and second peaks were reported by BOOMERanG (de Bernardis
\etal\ 2000).  These data could immediately be seen to be more consistent
with the no-CDM prediction than with the expectations of \LCDM\ as they
existed at the time (McGaugh 2000b).  The second
BOOMERanG data release (de Bernardis \etal\ 2002) measured
$\apk = 2.45 \pm 0.79$.  This is to be compared to the prior expectation
of $\apk < 2.06$ for \LCDM\ and $\apk \approx 2.40$ for no-CDM (Table~1).

\placetable{pkratios}

With the release of the WMAP first year data both \apk\ and \rpk\ are 
measured to high accuracy (Page \etal\ 2003).   
The WMAP data are in good agreement with the
predictions of McGaugh (1999b) for the case of no-CDM (Table~1).  They are,
in fact, bang on:  $\apk = 2.40$ (predicted) vs.\  $\apk = 2.34 \pm 0.09$
(measured), and $\rpk = 5.41$ (predicted) vs.\ $\rpk = 5.56 \pm 0.75$
(measured).  It is striking that the peak amplitude ratio is exactly that predicted
in advance by the no-CDM model.
  
\subsection{Constraints on the Baryon Density}

The parameter with the most leverage on the amplitude of the second peak 
is the baryon content.  Increasing the baryon fraction depresses the
amplitude of the second peak relative to the first, as the importance of
baryonic drag is increased relative to the forcing of the oscillations 
by CDM.  The observed peak amplitude ratio can thus be obtained by
decreasing the CDM density (to zero) at fixed baryon density, or by
increasing the baryon density while retaining CDM.  In order to be able to
distinguish between \LCDM\ and no-CDM, we must have a robust
independent constraint on the baryon density from BBN.

After the initial BOOMERanG measurements (de Bernardis \etal\ 2000) 
failed to detect the large second peak anticipated in \LCDM,
it was common to attribute this to a large baryon density 
($\ob = 0.031$: Tegmark \& Zaldarriaga 2000).  
This was considerably larger than implied by deuterium 
measurements at the time ($\ob = 0.019$: Tytler \etal\ 2000),
which were themselves surprisingly high compared to previous
BBN estimates ($\ob = 0.0125$: Walker \etal\ 1991).  Later, the 
baryon density implied by \LCDM\ analyses of the CBR data dropped
somewhat (de Bernardis \etal\ 2002).  
This was not because the second peak grew larger;
it merely resolved out at the amplitude predicted by no-CDM.
However, the lower limit on the amplitude of the peak was considerably
improved, excluding the very high baryon densities which would be
required to completely suppress the second peak.  That these had been
consistent with the earliest CBR data somewhat skewed our perspective.
After the second peak was clearly detected, the preferred baryon
density was still larger than the BBN value, but only by a small amount,
and not one which was significant once 
systematic errors (e.g., tilt) were considered (Netterfield \etal\ 2002).  

The WMAP best-fit model (Spergel \etal\ 2003) fits the small second peak 
with a baryon density of $\ob = 0.0224 \pm 0.0009$ and a tilt of
$n = 0.93$ with a large running of $n$ with scale.
Tilting the power spectrum in this manner
helps to suppress the second peak relative to the first.
The test I had proposed should maintain $n = 1$, which is perfectly
consistent with the WMAP
data themselves.\footnote{Running is inferred when other types
of data are included in the analysis, especially the power spectrum
from Lyman $\alpha$ clouds.}  In this case, the baryon density must 
be higher: $\ob = 0.024 \pm 0.001$.  It is interesting to compare these
to independent BBN constraints. 

\placetable{BBN}

Estimates of the baryon density \wb\ inferred from the light elements
$^2$H, $^4$He, and $^7$Li over the past decade are tabulated in Table~2
and plotted in Fig.~2.  BBN was already a very well developed field prior 
to 1995; earlier work is represented by the compilations of Walker \etal\ 
(1991) and Copi \etal\ (1995).  The baryon densities implied by analysis
of the CBR with CDM are also shown.  Without CDM the CBR is not a
precision baryometer, as the peak amplitude ratios are very similar for
the plausible range of baryon densities. (Fig.~1).  The WMAP peak data
are consistent with no-CDM models in the range $0.010 \le \wb \le 0.022$.

The BBN literature contains a wealth of information.  When authors have
quoted a value for \wb, it is given in Table~2.  Often, only
a measurement of the abundance of a particular element is reported.
In these cases, I have translated the result into
a baryon density using the calibration of
Burles \etal\ (1999).  For lithium, this determination can be 
double-valued; in such cases I have adopted the higher \wb.

Fig.~2 shows that the bulk of the data from all three elements fall 
in the range $0.005 < \wb < 0.02$.  Taking the data at face value,
both gaussian and biweight statistics give $<$\wb$>$$= 0.014 \pm 0.05$.
This value comes from the individual elemental abundances only,
and does not include previous compilations or CBR data.  This mean
is fairly robust to how the data are treated.  It does not change if extremal
points are rejected, nor if we accept only a single measurement from
each independent group.  

No-CDM models are consistent with all three of the light elements,
to the extent to which they are consistent with each other. 
In contrast, the baryon
density implied by \LCDM\ analyses of the CBR all have $\wb > 0.020$.
No measurement of any element ever implied a baryon density this
high until after the first CBR data appeared
(see also Steigman, Kneller, \& Zentner 2002).  So while it is true
that the WMAP \LCDM\ analysis (Spergel \etal\ 2003) 
is consistent with the latest deuterium estimate (Kirkman \etal\ 2003), 
the bulk of BBN data are more consistent with no-CDM.  

\subsection{CBR Data and Models: 2000 --- 2003}

Fig.~3 shows the impressive improvement of CBR data over the past
few years, together with the models advanced to explain them.
The top panels of Fig.~3 shows the initial data from the
BOOMERanG (de Bernardis \etal\ 2000) and Maxima-1 
(Hanany \etal\ 2000) experiments.  The BOOMERanG
data are shown both as published (open triangles) and
adjusted to agree with the Maxima-1 calibration
(filled triangles: McGaugh 2001).  On the left the $\wb = 0.019$
``\LCDM\ 1999'' model from Table~1 
is shown.  This case is very similar to the high
baryon fraction \LCDM\ model considered by McGaugh (1999b),
and is chosen for illustration as the {\it a priori\/} model which is
least inconsistent with the data.  A quite generic prediction
of \LCDM\ models had been 
a substantially larger second peak than was subsequently observed.

In contrast, the low amplitude of the second peak is quite natural
for the no-CDM model.  The $\Ob = 0.03$ model from McGaugh 
1999b is shown at right.  Parameters of the no-CDM models shown 
in Fig.~3 are given in Table 3.  Note that all of the no-CDM models 
shown in Fig.~3 existed before the data they are shown with.

The middle panel of Fig.~3 shows the second data release from 
BOOMERanG (de Bernardis \etal\ 2002) and the first data from 
DASI (Halverson \etal\ 2002).
On the left the best fit \LCDM\ model (with strong priors) of 
Netterfield \etal\ (2002) is shown.  This has changed substantially
from the prior expectation of \LCDM, yet still provides a worse
description of the data than does the no-CDM model (at right).
The model is only acceptable because of the systematic uncertainty
in the beam size, which allows for some play in the tilt of the power
spectrum.  

In contrast, the improved data have simply moved closer to the
no-CDM model as fit to the initial BOOMERang data release
(McGaugh 2000b) and normalized to Maxima-1 (McGaugh 2001). 
There is no need to iterate the fit at all.
Indeed, the revised calibration of BOOMERanG
was well predicted by the pre-existing no-CDM model.

The bottom panel shows these data and the first year WMAP data.  
The fit to the WMAP data (Spergel \etal\ 2003; left) has a larger \apk\ than 
the fit of Netterfield \etal\ (2002), matching that of the no-CDM prediction
(right).  The WMAP data are good enough to require the first 
tweak to the no-CDM fit, as the first peak in the WMAP data is
slightly narrower than indicated by previous data.  This small difference
has the interesting consequence of implying a significant neutrino mass.

\subsection{A No-CDM Model for the WMAP Data}

Providing a detailed fit to CBR data are beyond the ambition
of the no-CDM model.  We expect the new physics entailed by
MOND to render standard calculation of the temperature fluctuations
incorrect at some level.  The peak amplitude ratios are the most robust 
aspect of the no-CDM prediction since there is no obvious mechanism to
deviate from a pure damping spectrum in the absence of CDM.  

Nevertheless, pre-existing models do give a tolerable match to the detailed 
shape of the power spectrum (Fig.~3).  It is interesting to see if it is
possible to find a no-CDM model which matches the details of the WMAP data,
not just the peak amplitude ratio.   Such a model is shown in Fig.~4, with
details in Table~3.

\placetable{nocdmmod}

As a matter of principle, I restrict the free parameters of no-CDM as much as
possible with very strong priors.  For example, I fix the baryon density to the
value advocated by Tytler \etal\ (2000): $\wb = 0.019$.  Other baryon
densities are certainly viable, but the question here is not what baryon
density the CBR prefer.  Rather, we wish to know if there is a no-CDM model 
motivated by BBN which is consistent with WMAP.  Similarly, I fix 
$H_0 = 72\; \mathrm{km}\,\mathrm{s}^-1\,\mathrm{Mpc}^{-1}$
(Freedman \etal\ 2001) and $\tau = 0.17$ (Kogut \etal\ 2003).
Of course, the most important prior for a no-CDM
model is $\Omega_{\mathrm{CDM}} = 0$.

Three free parameters are required to fit the WMAP data.  The aspects of the
data which need to be matched are the position (in \LL) of the first peak,
the amplitude of the temperature fluctuations ($C_{\LL}$), and the width
of the peaks.  The positions of the peaks are controlled by geometry.
The amplitude of the fluctuations is somewhat arbitrary in CMBFAST models,
and can be addressed in several ways.  In order to match the width of the
peaks, it appears necessary to invoke massive neutrinos. 

It is perhaps too strong a term to describe what I have
done as a ``fit'' to the WMAP data.  In order to match the data, I have adjusted
three parameters:  \OL, $n_t$, and $f_{\nu}$.  \OL\ is adjusted to match the
location of the first peak.  A modest tensor component $n_t = 0.04$ is used to 
match the amplitude at small \LL.  The neutrino fraction $f_{\nu}$ is used 
to slim the peaks, which are a bit too fat in a purely baryonic models.
I have only adjusted these far enough to find a tolerable match to the
data.  I have not attempted an exhaustive exploration of parameter space;
modest improvements are no doubt possible.  However, I do not believe that
this would be a meaningful exercise since I expect MOND to cause real
deviations from a pure no-CDM model, especially at $\LL < 100$.
Each of these three parameters is discussed in turn below.

\subsubsection{Geometry}

The geometry of an FRW universe is well specified by the matter content
\Om\ and \OL.  A great success of the Inflationary paradigm is the apparent
flatness of the universe ($\Om+\OL \approx 1$) as indicated by the position
of the first acoustic peak $\LL_1 = 220$ (Page \etal\ 2003).  As noted
earlier, an oddity of a universe with a significant cosmological constant is
that while it is consistent with the Inflationary prediction of geometric flatness, 
it does not solve the coincidence problem which was one of the important
motivations for Inflation.  This is not important here; one can have either
or both of MOND and Inflation.  Neither requires the other, nor are they
mutually exclusive.

While the cosmic geometry of an FRW model is understood,
that in MOND is not.  So while a robust prediction for \apk\ can be made,
there is no prior expectation for the location of the first peak. 
This is treated as a parameter to be fit.  

There are two approaches which can be used to fit the peak location.
A model which is flat in the usual Robertson-Walker sense can be scaled
by a multiplicative factor $\alpha$ to match the peak position, so that
$\LL_1 = \alpha \LL_{\mathrm{model}}$ (McGaugh 2000b).  Alternatively,
one may treat \OL\ as a free parameter and find the value which places the
peak in the right location.  Either way, there is a single fit parameter.
I have adopted the latter approach here.

In effect, I am using the cosmological constant 
as a fudge factor to express our ignorance about the 
geometry in MOND.  This is not particularly different from the role it 
currently plays in standard cosmology.  As such, we should not invest
too much importance in the particular numerical value obtained.

One curious note is that while the geometry is {\it close\/} to flat, 
no-CDM models do require a slight but significant positive spatial
curvature.  Such a model is closed in standard cosmology, but the
significance of this fact in MOND is unclear.  Perhaps it contains a
clue to the nature of the underlying theory, or perhaps it is just a 
coincidence.  

The coincidence problem may be eased by the
proximity of the no-CDM geometry to the de Sitter solution, which is
an attractor in \Om-\OL\ space.  It is worth noting that in the absence
of a repulsive term ($\Lambda$), a MOND universe may eventually
recollapse for any \Om\ (Felten 1984).  Since $\Lambda$ is essentially 
just a fudge factor encapsulating our ignorance of the geometry, it remains
conceivable that we live in such a Felten universe where there is no 
critical value of \Om, and hence no coincidence problem.

\subsubsection{Amplitude}

As with geometry, the amplitude of a model can be scaled by a multiplicative
factor to match the data:  $C_{\LL} = A C_{\LL,\mathrm{model}}$.
There are many uncertainties in the normalization of CBR models, so
this is fair within plausible bounds.  This can be seen in the various
models presented by Spergel \etal\ (2003).

Various factors affect the normalization, including the optical depth and
neutrino mass.  If we get these things right, we should be able to get the
normalization right ($A = 1$).  Traditionally, CBR models have been
normalized to COBE at $\LL = 10$.  This is not the best choice for a
no-CDM model, as one of the predicted post-recombination effects of
MOND is an enhancement of the ISW effect (McGaugh 1999b).  
This should be most pronounced on large scales.

The ISW effect arises from the 
variation of metric fluctuations during structure formation.
Unfortunately, it is not presently possible to make an explicit calculation
of this effect in MOND.  However, we do expect MOND to cause
rapid structure formation. The rapid variation of the effective
potential should enhance the ISW.  This will distort the shape of the
CBR power spectrum so that there is an excess of power at $\LL < 100$
over that in a pure no-CDM model which is normalized to the amplitude
of the peaks (at $\LL \approx 220$).

The pre-existing no-CDM model shown in the bottom panel of Fig.~3
fits the WMAP data well at $\LL > 100$.  
It under-predicts the observed power at $\LL < 100$ (Fig.~5).
This is, qualitatively at least, the expected signature of the ISW in MOND.
However, it is interesting to see if we can obtain a fit to all of the data.
In order to make up the power on large scales, we invoke a modest
tensor contribution.  This is the only significant difference between the
fit in Fig.~4 and the old no-CDM model with neutrinos.

The tensor contribution is the standard inflationary complement to a mildly
tilted spectrum:  $n_t = 1-n = 0.04$.  Such a tensor contribution
would be consistent with a broad range of inflationary models (e.g.,
Tegmark \etal\ 2003).  As Inflation occurs very early, and the effects of MOND
are only manifest after recombination, it is quite possible to have both.

While it is conceivable that there literally is
a tensor component, really this is just a fudge to show that it is
possible to fit the low-\LL\ data with a no-CDM model. 
For the reasons described
above, we do not expect to be able to fit a no-CDM MOND proxy model to
$\LL \lesssim 100$.  Instead, the deviation at
small \LL\ illustrated in Fig.~5 should be viewed as an empirical constraint
on the ISW in MOND.  How well low redshift structure should correlate with 
the temperature fluctuations in the CBR is unclear because of the highly
non-linear nature of MOND.  But one would naively expect a strong signal
in this scenario.

An interesting consequence of this approach is that further scaling (by $A$)
is hardly necessary.  That is, $n_t$ not only fits the low-\LL\ data, but also
fixes the overall normalization to well within the uncertainty of the optical
depth (fixed to $\tau = 0.17$: Kogut \etal\ 2003).  It would perhaps be
better to normalize no-CDM models at $\LL = 220$ than at $\LL = 10$.

\subsubsection{Neutrino Mass}

One peculiar aspect of the fit shown in Fig.~3 is a high neutrino fraction,
$\fnu \approx 0.65$.  In order to obtain a detailed
no-CDM fit to the WMAP first year data, it is helpful to have
heavy neutrinos.  The CBR parameter space is very large, so it is difficult to
say whether these are required.  However, models with neutrino densities
comparable to or slightly in excess of the baryon density do seem to be
preferred.

The need for heavy neutrinos is brought on entirely by the narrowness of
the first peak.  This is rather subtle, being the difference between the 
models in the bottom right panel of Fig.~3.  While McGaugh (1999b) 
noted that such an effect was
possible, it is only perceptible in data of WMAP quality.  

The model in Fig.~3 has three 
degenerate neutrinos with $\mnu = 1.1$~eV.  The neutrino 
mass\footnote{The inferred neutrino mass is also very sensitive to the
physics of recombination, changing upwards 20\% between CMBFAST
v4.0 and 4.3.} is not well determined.  One issue is a degeneracy between
neutrino mass and optical depth.  Both affect the absolute amplitude of the
observed temperature fluctuations in similar ways.  If the optical depth
were found to be larger than 0.17, the neutrino mass would go down.
Another effect is due to the baryon density.  The strongest constraint is on
the neutrino fraction; the neutrino mass scales roughly with the baryon
density.  Lower baryon densities are tolerable, so lower neutrino masses
would be as well.

There are other indications of a finite neutrino mass in MOND.
Perhaps the most significant is the residual mass discrepancy 
in clusters of galaxies (Aguirre \etal\ 2001; Sanders 2003).
A neutrino mass of 1 to 2 eV would be about right to explain this.
Neutrinos of this mass would not be trapped by galaxy scale potentials,
but would be by clusters, and would provide about the right amount of mass.

While the neutrino mass is not well constrained at present,
it is subject to independent laboratory tests.  The prospects for measuring
a mass as high as $\mnu \sim 1$~eV in upcoming experiments are good;
there is already a claim that $\mnu \approx 0.39$ albeit with a large
uncertainty (Klapdor-Kleingrothaus \etal\ 2001).
Such a measurement would give us another means of distinguishing between
scenarios.  Structure can only form in \LCDM\ if $\mnu < 0.23$~eV
(Spergel \etal\ 2003).  A firm laboratory measurement of a neutrino mass
heavier than this would constitute a clear falsification of \LCDM. 
The rapid structure formation in MOND is not so adversely affected
to this limit, and some power suppression from heavy neutrinos might even be 
desirable to prevent MOND from overproducing structure
(Nusser 2002; Knebe \& Gibson 2003).

Presumably, if MOND is correct, it is only the approximate expression of
more general theory from which the proper cosmic geometry should be
derived.  By that token, the Robertson-Walker geometry would also be
an approximation of the deeper theory.  We should not expect that one
fudge factor ($\Lambda$) can necessarily connect both.  
It is conceivable that the approximation to the geometry we are making
is inadequate for simultaneously
matching the locations and widths of the peaks.  These are extremely 
sensitive to the geometry.  It is therefore difficult
to exclude the possibility that a deeper, relativistic theory
could fit the observations without heavy neutrinos.  
However, they do seem to be the most obvious way to explain
the modest width of the acoustic peaks without CDM. 

\subsection{The Third Peak}

The WMAP first year data drop to $S/N < 1$ for $\LL > 660$
(Bennett \etal\ 2003), so do not, as yet, provide any useful constraint on the
amplitude of the third peak (Page \etal\ 2003).  
Fig.~6 shows the data from WMAP together with other experiments which
probe smaller angular scales:  CBI (Mason \etal\ 2003),
ACBAR\footnote{Data obtained from 
http://cosmology.berkeley.edu/group/swlh/acbar/data/index.html.}
(Kuo \etal\ 2003), and VSA (Grainge \etal\ 2003).
The other experiments
provide some hints, but do not yet define the third peak to the
quality required for a clear test. 

A critical issue with mixing data from different experiments is the mutual
consistency of their calibrations.  This is an ancient problem in astronomy
from which CBR experiments are not immune.  Indeed, a prominent
example of the problems faced in comparing such data has already been 
provided in the first data released by BOOMERanG and MAXIMA-1.  
Even though the {\it shape\/} of the power spectrum measured by the
two experiments has always been consistent, their absolute calibration
was clearly offset in the first data release.  There is little reason to expect
that the calibrations of all the independent experiments are in perfect
accord.

It is possible to place the various experiments on a mutually consistent
scale by normalizing them to a common reference.  Once a reference
calibration is adopted, the other experiments can be brought onto a
common scale by a multiplicative factor ${\cal F}$ which maps
$C_{\LL} \rightarrow {\cal F} C_{\LL}$.  ${\cal F}$ is determined by
minimizing $\chi^2$ for the difference in $C_{\LL}$ between each 
experiment and the reference over the range of \LL\ where they overlap.

This method was employed by McGaugh (2001) to bring the first
BOOMERanG data into accord with the MAXIMA-1 calibration.
The result was successful in anticipating the subsequent recalibration
of BOOMERanG (Fig.~3).  The obvious choice of reference calibration
now is WMAP.  The scaling ${\cal F}$ determined relative to WMAP for 
each experiment is given in Table~4.  For the most part, these imply
that the normalizations of the experiments which probe high-\LL\
need to be reduced somewhat to be consistent with WMAP.  This
can be seen by eye in the top panel of Fig.~6, where the CBI and
ACBAR data fall above the WMAP data in the range of \LL\ where
they overlap.  This was also noted by \"Odman (2003), who pointed out that
there is some tension between low-\LL\ and high-\LL\ experiments.

\placetable{fudgefact}

After the initial submission of this paper, new data were reported by
CBI (Readhead \etal\ 2004) and VSA (Dickinson \etal\ 2004).
These are shown in the middle panel of Fig.~6.
The new results included an explicit evaluation of the calibration difference
of these experiments with respect to WMAP based on their observations
of Jupiter.  This allows us to compute the effective ``observed'' correction
factor ${\cal F}_{obs}$ as a check on the previously computed ${\cal F}$.
In the case of the VSA, as with BOOMERanG previously, the two are
consistent.  In the case of CBI, ${\cal F}$ is somewhat less than 
${\cal F}_{obs}$.  This appears simply to be due to the large uncertainty in
${\cal F}$ in this one case:  the initial CBI data only have two independent
points overlapping the range of \LL\ constrained by WMAP, and one of those
has a very large error bar.  The procedure of scaling by ${\cal F}$ does
therefore appear to be valid.

The reduced $\chi_{\nu}^2$ of the no-CDM model with respect to WMAP and
the other CBR experiments are given in Table~4.  The match to the WMAP TT 
data is treated as predictive of what the other experiments should observe:
no further adjustment is made to the model.  A plot of the $\chi^2$ values
of the individual data points is given in Fig.~7.

A few badly fitting points stand out.  The single most deviant point
is that of WMAP at $\LL = 40$.  Like the point at $\LL = 210$ which gives
a notched appearance to the first peak, this point at $\LL = 40$ is deviant from
any smooth model.  Since the use of a tensor contribution to mimic an excess
ISW effect from MOND is far from perfect, this is not of great concern.
To show the impact of this and the other stand-out points on the $\chi^2$
budget, Table~4 also gives $\chi_{\nu,mod}^2$ without them.

A deviant point of greater concern is that of the VSA at $\LL = 795$.
This indicates a third peak significantly larger than predicted by the no-CDM
model.  The absolute value of $C_{\LL}$ of this point dropped in 2004 from
its previous value in 2003, but the error bar also shrank considerably,
making the statistical significance of the deviation greater.  Still,
the overall fit with respect to this experiment is not terrible, and it
is rather odd that the lion's share of the $\chi^2$ budget is due to
this single point.

The situation is similar with respect to BOOMERanG.  The $\chi^2$ budget is
dominated by the point at $\LL = 700$.  The overall fit to this experiment is 
nevertheless acceptable.  Indeed, my $\chi^2$ do not account for the tilt in
the power spectrum allowed by the systematic uncertainty in the beam size 
(Netterfield \etal\ 2002), so would come down somewhat if this were taken into
consideration.  The no-CDM model is clearly acceptable to the empirical
peak location estimates of de Bernardis \etal\ (2002) (top panel of Fig.~6).

The fit with respect to Maxima (Lee \etal\ 2001) and Archeops
(Benoit \etal\ 2003) is acceptable without caveats.  That with respect
to DASI is rather poor.  This is in part because of
a hint of a large third peak in those data, but roughly half the
$\chi^2$ budget is contributed by a point at small \LL\ where the WMAP data
have been given precedence.  This would seem to tell us more about the 
difficulty of error estimation in real experiments than about the model.

The worst $\chi^2_{\nu}$ is that of ACBAR, with the calibration as published.
However, if we scale the ACBAR data by ${\cal F} = 0.78$ to match the WMAP
data in the range where they overlap, $\chi^2_{\nu,mod}$ is quite acceptable.
Indeed, once this scaling factor is accounted for, the no-CDM model does an 
excellent job of predicting the undulations of the ACBAR data, which appear
to hit the fourth and fifth peaks and their preceding troughs.  Whether these
data indicate a large or small third peak hinges entirely on the true value of
${\cal F}$.  There are four points with tolerable error bars overlapping the 
WMAP data, so ${\cal F}$ is better determined than for CBI, but not as well
as for the other experiments.  

Multiple experiments measure power in the vicinity of the third peak
in excess of that predicted by the no-CDM model.  The agreement between
independent experiments is encouraging, but the data are not yet 
adequate to clearly reject this aspect of the no-CDM model.  One must
weigh the merits of all the predictions together; as yet the amplitude
of the third peak remains the least well constrained of the tests discussed
here.  To be really sure, we would like to see the first three peaks all
measured by the same experiment with high quality so that the third peak
is at least as well defined as the second one is in the WMAP first year data.

\section{Signatures of MOND-like Physics}

While the no-CDM {\it ansatz\/} is a good first approximation to what we
might expect for the CBR in MOND, it must fail at some level.  We have known
for a long time that the growth factor between the epoch of recombination
and the present epoch is too large to be explained by gravitational collapse
with normal gravity and the observed baryons.
This does not in itself require CDM: one could also imagine 
a change to the force-law which results in a faster growth rate.

Recently, the formation of structure under MOND has been considered
by a number of authors (Sanders 1998, 2001; McGaugh 1999c;
Nusser 2002; Stachniewicz \& Kutschera 2002; Knebe \& Gibson 2003).
A variety of assumptions and approximations have been made in these
works, but one generic result is that MOND causes structure to form
fast.  The onset of
structure formation is necessarily delayed until after recombination
(there is no CDM component immune to radiation pressure which
can clump up earlier) and matter domination (which occurs after
recombination for $\Om = \Ob$).  Once these conditions are met and
the perturbations enter the MOND regime, the non-linear MOND force law
causes structure formation to proceed rapidly.  Bright
($\sim L^{\star}$) galaxies can be in place by $z \sim 10$;
clusters by $z \sim 3$ (Sanders 1998).  The growth of structure slows
or even saturates in accelerating universes (Sanders 2000), so the 
resulting picture is one of a very quiescent early universe undergoing
a dramatic period of rapid structure formation which then eases off
(see Fig.~1 of Sanders 2000).
This contrasts sharply with the steady, gradual build up
of structure in CDM models, and should leave subtle but recognizable
signatures in the CBR.

\subsection{Early Reionization}

Perhaps the most obvious signature of MOND-induced structure formation
is an early onset of reionization.  In CDM models, no significant mass
should form in stars before $z \sim 7$ (e.g., White \& Frenk 1991;
Stachniewicz \& Kutschera 2003).
Consequently, reionization of the universe was expected to occur
fairly late (Loeb \& Barkana 2001), giving little opportunity for
scattering of the CBR by free electrons.  This translates to the expectation
that there should be little polarization of the CBR and low optical depth to
the surface of last scattering.  The situation shortly before the WMAP data
release was summed up by Peacock (2003):
``For reionization at redshift 8, we would have $\tau \approx 0.05$;
it is unlikely that $\tau$ can be hugely larger.''

Kogut \etal\ (2003) report a surprisingly large polarization signal of
$\tau = 0.17 \pm 0.04$.  This pushes the epoch of reionization
towards $z \approx 17$,
much earlier than nominally expected in \LCDM.  While the difference
between $\tau = 0.05$ and 0.17 may not sound large, bear
in mind the inverse relation between time and redshift.  The universe is
$\sim 500$ Myr old at $z = 8$; one of the advantages of \LCDM\ is that
it forms structure this ``early'' (Mo \& Fukugita 1996). 
At $z = 17$ the universe is only 180 Myr old, a very early time by the
standard of any prior expectation in the context of CDM. 
The observed optical depth {\it is\/} hugely larger than expected. 

In contrast, early structure formation is a natural consequence of MOND.
There is no need to invoke super-massive Pop.~III stars as is currently
popular in \LCDM\ models.  Stars with a normal IMF simply form sooner.
This distinction was noted by McGaugh (1999b), where it was predicted that
the polarization signal would be higher than nominally expected in \LCDM.

While the qualitative prediction of early reionization is clearly realized
in the WMAP data, it is rather more difficult to estimate a specific number
for the optical depth.  This depends not only on the timing of structure
formation, but also on details of the star formation process and the liberation
of ionizing photons which are
uncertain in either case.  Still, we can make a crude estimate based on
the work of Sanders (1998, 2000) and Stachniewicz \& Kutschera (2002).  
Roughly speaking, $L^{\star}$ galaxies may be in place by $z \approx 10$,
with smaller (globular cluster) scale lumps forming even earlier, perhaps as
early as $z \approx 100$ (Stachniewicz \& Kutschera 2002).  Stars take a
finite amount of time to form, and
it remains uncertain how long it takes to form a significant number.
However, it seems quite plausible for reionization to have occurred
at $z \gtrsim 15$.  MOND is therefore quite consistent with
$\tau \gtrsim 0.15$, and might conceivably produce even larger optical depths 
(e.g., $\tau \approx 0.3$:  Fig.~8).  
This fits nicely with the reionization epoch determined
from the cumulative mass function of Lyman $\alpha$ systems, 
$z = 24 \pm 4$ (Popa, Burigana, \& Mandolesi 2004).  It might also help
to explain the excess power at $\LL > 2000$ reported by CBI (Bond \etal\ 
2002), as more clusters will be in place earlier than in \LCDM, enhancing
the SZ effect.

\subsubsection{Are Super-Massive Pop.~III Stars Viable?}

If the optical depth is really as large as 17\%, reionization must occur early
and structure must form fast.  One way to achieve this in \LCDM\ is to increase
the amplitude of the power spectrum on small scales.  However, this would
contradict the limits on power in the fluctuation spectrum on small scales at
$z = 0$ (Bullock \& Zetner 2002; McGaugh, Barker, \& de Blok 2003;
Somerville, Bullock, \& Livio 2003; van den Bosch, Mo, \& Yang 2003).

Ordinary stars and quasars will not suffice to produce the observed optical
depth within the limits on the power spectrum.
It is necessary to invoke super-massive Pop.~III stars, or some other object
which is very efficient at producing ionizing radiation.
The earliest objects to collapse must be $\sim 50$ times as efficient
at converting mass to ionizing photons as are collapsed objects at the present
time (Sokasian \etal\ 2003a).  

Obtaining the required ionizing flux early enough is quite a stretch.
Fukugita \& Kawasaki (2003) argue that the earliest reionization can occur
is $z \approx 13.5$, with an upper limit on the optical depth of
$\tau < 0.17$ --- right at the observed value.  Ricotti \& Ostriker (2003)
find that $\tau > 0.12$ can only be achieved with great contrivance even
with Pop.~III.  One requisite detail is a remarkably high UV escape
fraction (Sokasian \etal\ 2003b).  These considerations make one wonder
if it is reasonable to invoke super-massive Pop.~III stars.

Pop.~III stars have long been sought as
the first generation of stars with primordial abundances.  
While very metal poor stars have been discovered
(e.g., Norris, Beers, \& Ryan 2000), no examples of
metal free stars have ever been found.  At [Fe/H] $= -3.7$, the star
discussed by Norris \etal\ (2000) is consistent with enrichment by the
supernova of a single metal free $\sim 30 M_{\sun}$ star 
(Woosley \& Weaver 1995).  Since the formation time for low mass
stars exceeds the lifetime of massive ones, it seems plausible that 
long-lived stars of the first generation were enriched by the massive
stars of the first generation to a sufficient
degree that no metal-free stars remain today.

A $\sim 30 M_{\sun}$ star is perfectly normal, and will not suffice to
achieve the early reionization of the universe.  Metal free stars have less
opacity than Pop.~I stars and should therefore be more efficient producers
of UV radiation, but the net gain due to this effect is only a factor
of $\sim 3$ (Schaerer 2003).  To obtain the desired factor of $\sim 50$
increase, Pop.~III must have had a peculiar IMF which produced many
super-massive (200 --- $500\,M_{\sun}$) stars.

It is commonly thought reasonable that Pop.~III have an IMF skewed
towards massive stars.  The line of argument is that since Pop.~III
stars form from meal-free material, gas cooling is less efficient, leading to
a larger Jeans mass and hence a larger characteristic stellar mass.  At this
hand--waving level, super-massive Pop.~III stars seem fairly natural.
If we proceed with the logic of this argument, the IMF should be a
strong function of metallicity.

A good deal is known about stars.  As noted in the review by
Hillenbrand (2003), ``there is little evidence for substantial 
variations in the IMF from region to region, or over time.'' 
The IMF appears to be universal, with no evidence for the expected
dependence on metallicity (Kroupa 2002).  This
includes the IMF of massive stars (Oey, King, \& Parker 2003).
There is no indication that the early-time IMF was different from
what it is now, either from direct counts of stars in ancient, 
low metallicity systems (Wyse \etal\ 2002), or from
the integrated cosmic star formation history
(Baldry \& Glazebrook 2003).  Moreover, dynamical constraints,
especially the small scatter in the Tully-Fisher relation, allow very
little room to consider a variable IMF (McGaugh \etal\ 2000;
Bell \& de Jong 2001; Verheijen 2001; McGaugh 2004), at least when
averaged over the scales of galaxies.  The hypothesis that
metallicity affects the IMF can be tested directly in
low metallicity star forming regions like the
Magellanic clouds, where again there is no evidence for a systematic
variation of the IMF (Massey 2002).  There {\it is\/} good evidence that
the UV spectral hardness increases with declining metallicity, 
but this is entirely explained by the lower opacity of lower metallicity 
stars with the same IMF (McGaugh 1991).  

A considerable and diverse body of evidence disfavors the notion that the 
IMF varies.  From what we know empirically about stars over a broad 
range of metallicity, it seems that the Jeans scale does not play a strong
role in determining the IMF.  From this perspective,
there is no reason to suppose that Pop.~III stars would have
formed with a radically top-heavy IMF.  One can
of course imagine some discontinuity at
zero metallicity, and there are certainly reasons why such a
discontinuity might occur (e.g., Bromm \& Loeb 2003).
However, this is no guarantee that such stars would form with the
required properties.  Super-massive  
Pop.~III stars are only inferred to exist in order to cause what for \LCDM\ 
is abnormally early reionization.  This is not sufficient
evidence to prove their existence.

In contrast, early structure formation is natural to MOND.  
Normal stars suffice; they simply form earlier. 
There is no need to invoke a population of super-massive Pop.~III stars.  
It remains to be seen whether substantial improvements to the polarization
measurement of Kogut \etal\ (2003) can be made, or if any such refinements
would lead to grossly different interpretations.  A drop in the optical depth
to $\tau < 0.12$ would favor CDM, while $\tau > 0.12$ favors MOND.

\section{Further Tests}

Recent measurements of temperature fluctuations in the CBR are a vast 
advance over the situation of only a few years ago.  However, they are not
yet sufficient to distinguish between the very different world models of
\LCDM\ and MOND.  On the one hand, the \LCDM\ models discussed
by Spergel \etal\ (2003) provide an excellent fit to the WMAP data.  On the
other hand, the second peak is much smaller than it was predicted to be
by \LCDM\ prior to its measurement, while being exactly the amplitude
which was predicted by McGaugh (1999b) for the case of no CDM.  In
addition, the surprisingly large polarization signal measured by WMAP
(Kogut \etal\ 2003) is also expected from early structure formation with
MOND.  The confirmation of these two {\it a priori\/} predictions of
McGaugh (1999b) would be a remarkable coincidence if \LCDM\ were
correct.  

One might think it also a remarkable coincidence that \LCDM\ fits the WMAP
data so well if it is incorrect.  This is not really the case.  Until much
better constraints on the third peak become available, one expects to be able 
to fit a \LCDM\ model to the data even if MOND is the correct underlying 
physics.  Indeed, because of the deviation from a pure no-CDM model
expected from early structure formation in MOND, conventional fits 
{\it should\/} imply the need for some CDM.  

There are other observations which might help to distinguish between
\LCDM\ and MOND.  A few of these are explored in this section.  This is
by no means a comprehensive list.  It is merely intended to give expectation
values which can be tested against new data.

\subsection{Baryonic Features in the Galaxy Power Spectrum}

The acoustic oscillations observed in the CBR will be frozen into the power
spectrum if not washed out by dark matter.  This predicts (McGaugh 1999c;
Sanders 2001) that there will be sharp spikes in the galaxy power spectrum
$P(k)$ at low redshift.  On small scales, these may well be wiped out by the
highly nonlinear growth of structure in MOND, which may lead to the 
exchange of power between wavenumbers (it can not be assumed that
individual $k$ will remain independent).  Indeed, Nusser (2002)
finds convergence to a unique slope of the power spectrum irrespective
of initial conditions.  An overall shape for the power spectrum
similar to that of \LCDM\ is found by Sanders (2001), with a bit more power 
on large scales relative to small ones.

There is some hope that the baryonic features on the largest scales will
survive the nonlinear evolution.  If so, they will appear as sharp, narrow,
down-going spikes in a power spectrum with an otherwise smooth 
envelope (McGaugh 1999c;
Sanders 2001).  Tegmark \etal\ (2003) note that similar bumps and wiggles
are present in three independent data sets: PSCz, 2dFGRS, and SDSS (see
their Fig.~36).  These are suggestive of the predicted baryonic features,
but are hardly convincing detections.  

Detecting the narrow spikes predicted as baryonic features is extremely 
challenging experimentally.
It requires not only large scale surveys, but also very fine resolution
in $k$-space.  The window functions typically used in power spectrum
analyses are broader than the predicted features, and will act to wash out
their appearance if present.  The experiment may be easier to perform at
high redshift (Blake \& Glazebrook 2003), though the rapid growth of
structure in MOND may mitigate against this.

Nevertheless, the presence of sharp, strong baryonic features provides
another means to discriminate between CDM and MOND.  $P(k)$ should
be smooth and featureless in CDM.  In MOND it should contain sharply
pronounced features on large scales.  

\subsection{Structures at High Redshift}

Another observation which may help to distinguish between CDM and MOND
is the amount of structure at high redshift.  As noted by 
Sanders (1998; 2001) and McGaugh (1999c), structure should form
considerably more rapidly in MOND than in CDM.  There are many
indications of early structure formation, and indeed, this was one of the
many indicators which pushed us towards \LCDM.  The presence of
a finite cosmological constant provides more time to form structure early
relative to previously conventional CDM models (e.g., Mo \& Fukugita 1996).

Early structure formation is quite natural in MOND (Sanders 1998, 2001;
McGaugh 1999c; Nusser 2002; Stachniewicz \& Kutschera 2002), so
having structure in place at high redshift is quite reasonable.  Indeed,
the problem MOND suffers may be overproducing structure (Sanders 1998;
Nusser 2002; Knebe \& Gibson 2003), perhaps by a factor of $\sim 2$.
This is sensitive to a variety of assumptions and the subject is not yet well
developed (Sanders \& McGaugh 2002), so it is unclear how serious a problem
this is.  Moreover, it may be mitigated if neutrinos turn out to be heavy. 
Late times are harder to predict in both CDM and MOND, so the strongest
test is at high redshift.

The evidence for structure at high-$z$ has grown considerably in recent
years.  Lyman break galaxies at $z \sim 3$ are already highly clustered
(Steidel \etal\ 2000, 2003), leading to the inference that the are very highly 
biased ($b \sim 6$ as opposed to $\sim 1$ for galaxies of comparable mass
at $z = 0$), or not massive at all, with luminosities vastly enhanced by
collisional star bursts (Somerville, Primack, \& Faber 2001).  In MOND, 
these would simply be the early stages of normal galaxies.  
Chen \etal\ (2003) find that massive galaxies are surprisingly abundant and
metal rich at $z > 1$.  Rocca-Volmerange
\etal\ (2003) find that large ($M_{\star} \approx 10^{12} M_{\sun}$)
galaxies are already in place at $z \approx 4$, much earlier than should be
the case with CDM.  $L^{\star}$ galaxies can form as early as $z \approx 10$
in MOND (Sanders 1998), so these observations present no puzzle.
Not only should massive objects form early, but they should also form into
large scale structures early.  Palunas \etal\ (2003) present evidence for large 
voids being present in the galaxy distribution already at $z \approx 3$, for 
which they find only a $\sim 1\%$ probability in \LCDM.

The high redshift universe appears to be rich in structure, perhaps 
surprisingly so for our expectations from CDM.  Such early
structure formation is quite natural in MOND.  If the latter is correct,
we would expect to continue to find large numbers of massive galaxies
already in place at $z > 3$, and rich clusters of galaxies at $z > 1$.
Of course, the stellar content of these systems must evolve in either
case, and mass evolution from merging may also occur in both.
But the expectation is for there to be more structure sooner
in MOND than one would nominally expect with CDM.

\section{Conclusions}

McGaugh (1999b) predicted various features of the power spectrum of
temperature fluctuations in the CBR for no-CDM
universes inspired by MOND.  These predict
\begin{itemize}
\item a specific first-to-second peak amplitude ratio, $\apk = 2.4$; and
\item a third peak smaller than the second.
\end{itemize}
These stem from the stipulation that there should be no CDM in a MOND
universe so that baryonic damping dominates the power spectrum.  The 
specific value of \apk\ follows from the baryon density as given by BBN,
with very little variation for plausible values of \ob.  Beyond
the no-CDM model, one does expect some signature of MOND in the CBR.
MOND-specific predictions are
\begin{itemize}
\item an early epoch of reionization leading to a large polarization 
signal; and
\item an excess of power on large ($\LL < 100$) scales.
\end{itemize}
I have tested these predictions with the WMAP first 
year data and other recent CBR experiments.

WMAP provides tests of all of these predictions except the third peak, 
to which its data do not yet extend
($S/N < 1$ for $\LL > 660$: Bennett \etal\ 2003).  
For the three items where WMAP alone provides a strong test, all of the MOND
predictions are confirmed.  Other data must be invoked to test the third
peak.  These do not confirm the no-CDM prediction, but do not clearly
reject it (\S 2.5).

The first-to-second peak amplitude ratio measured by WMAP
is $\apk = 2.34 \pm 0.09$ (Page \etal\ 2003).  The amplitude
of the second peak is considerably smaller than had been anticipated 
by \LCDM\ models prior to observational constraints on its value.  
Yet it is accurately predicted with no free parameters by no-CDM
(McGaugh 1999b).  It follows simply from setting
$\Omega_{\mathrm{CDM}} = 0$ and \ob\ to its BBN value. 

This peak ratio \apk\ is very sensitive to the baryon content of the universe. 
The range of baryon densities acceptable to no-CDM models is in
good agreement with BBN constraints from $^2$H, $^4$He, and $^7$Li.
\LCDM\ requires a significantly higher \ob\ than was indicated
by BBN prior to measurement of the second peak (\S 2.2).
Our plain ``vanilla'' \LCDM\ has been spiked with extra baryons.

The large polarization signal reported by WMAP (Kogut \etal\ 2003) is
surprising in the context of \LCDM, but not in MOND.  MOND is expected to 
grow structure much more rapidly than the conventional $\delta \sim t^{2/3}$.
Consequently, the first stars can readily form by $z \gtrsim 15$. 
The high optical depth reported by WMAP confirms this prediction. 
There is no need to invoke super-massive Pop.~III stars. 
A review of constraints on the sensitivity of the IMF to metallicity
suggests that these are rather contrived (\S 3.1.1).

Early reionization is one clear indication of MONDian physics beyond the
simple no-CDM model used to predict the peak amplitude ratios.
Another possible signature of MOND is the excess of power at $\LL < 100$
over the nominal no-CDM prediction.  This may be due to a strong ISW
effect as anticipated by McGaugh (1999b).  Though the data are qualitatively
consistent with this prediction, a quantitative computation of this effect in
MOND remains lacking.  

Obtaining a detailed fit to the WMAP data (not just the peak ratio) requires
three fit parameters: one for the position of the first peak, one for its
width, and one for the amplitude.  One combination of parameters that works
is $\OL = 0.92$, $\fnu = 0.65$, and a tensor component $n_t = 0.04$. 
These should not be taken too literally, as they
may simply encapsulate our ignorance of the underlying theory.  The apparent
tensor contribution in particular is simply a fudge to fit the data
at $\LL < 100$ where a deviation from the standard calculation is
expected from an enhanced ISW effect.

One interesting aspect of the detailed no-CDM fit is the apparent
need for a massive neutrino ($\mnu \sim 1$ eV).  There are a variety of
effects which might cause this inference to be incorrect (see discussion in
\S 2.4.3).  While a neutrino mass this high may seem unappealing, it provides
an additional means of distinguishing between CDM and MOND.  \LCDM\ 
models can not form sufficient structure if $\mnu > 0.23$ eV 
(Spergel \etal\ 2003),
so a robust laboratory measurement in excess of this value would in
principle falsify CDM.  While a smaller neutrino mass would not exclude
MOND, it would make it difficult to understand rich clusters of galaxies
(Aguirre \etal\ 2001; Sanders 2003). 

Three of the four predictions listed above have been realized in the WMAP 
data.  The prediction of a low third peak does not appear to be realized
in other data set.  If improved measurements confirm a second-to-third
peak amplitude ratio $A_{2:3} \lesssim 1$, it would favor CDM. 
In this case, the success of the other MOND predictions may just be a 
remarkable fluke.  On the other hand, a definitive
measurement of $A_{2:3} > 1.5$
would provide the clearest possible falsification of the existence of
non-baryonic cold dark matter.

\acknowledgements 
I would also like to thank Bob Sanders for his encouragement 
and persistent questioning on these issues.  I am also grateful to
the referee for suggestions which improved the organization of this text.
The author acknowledges the support of the University of Maryland 
through a semester General Research Board award.  
The work of SSM is supported in part by NSF grant AST0206078
and NASA grant NAG513108.


\begin{deluxetable}{lccc}
\tablewidth{0pt}
\tablecaption{Peak Ratios\label{pkratios}}
\tablehead{ & \colhead{$\omega_b$} & \colhead{\apk} & \colhead{\rpk} }
\startdata
\LCDM\tablenotemark{a} & 0.019\phn & 1.95\phn & 3.79 \\
~(1999) & $< 0.0214$\phn\phn & $< 2.06\phn\phn\phn $ & $< 4.05\phn\phn $ \\
No CDM\tablenotemark{a} & 0.0056 & 2.57\phn & 7.61 \\
 & 0.011\phn & 2.37\phn & 5.72 \\
 & 0.017\phn & 2.40\phn & 5.41 \\
WMAP\tablenotemark{b} & \dots & 2.34\phn & 5.56 \\
\enddata
\tablenotetext{a}{Values expected prior to any observational constraints (1999).
\LCDM\ models are optimistic in that they represent the largest values of the
ratios considered plausible at the time.  The three no-CDM models are the
$\Omega_b = 0.01$, 0.02, and 0.03 predictions from McGaugh (1999b).}
\tablenotetext{b}{Page \etal\ (2003).  The uncertainty in \apk\ is $\pm 0.09$
and that in \rpk\ is $\pm 0.75$.}
\end{deluxetable}

\clearpage

\begin{deluxetable}{lccl}
\tablewidth{0pt}
\tablecaption{Baryon Density\label{BBN}}
\tablehead{ 
\colhead{Measurement} & \colhead{\ob\tablenotemark{a}}
 & \colhead{$\pm \sigma$} & \colhead{Reference} }
\startdata
Compilation 
& 0.0125 & 0.0025  & Walker \etal\ (1991) \\
& 0.0145 & 0.0055  & Copi \etal\ (1995)\tablenotemark{b} \\
& 0.0066 & 0.0010  & Fields \etal\ (1996) \\
& 0.0190 & 0.0012  & Tytler \etal\ (2000) \\
$^2$H 
& 0.006\phn  & 0.001\phn   & Rugers \& Hogan (1996) \\
& 0.0193 & 0.0014  & Burles \& Tytler (1998a) \\
& 0.019\phn  & 0.001\phn   & Burles \& Tytler (1998b) \\
& 0.0188 & 0.0010  & Burles \etal\ (1999) \\
& 0.025\phn  & 0.001\phn   & Pettini \& Bowen (2001) \\
& 0.0205 & 0.0018  & O'Meara \etal\ (2001) \\
& 0.0214 & 0.0020  & Kirkman \etal\ (2003) \\
$^4$He 
& 0.0058 & 0.0012  & Olive \& Steigman (1995) \\
& 0.0125 & 0.0025  & Izotov \etal\ (1997) \\
& 0.0066 & 0.0010  & Olive \etal\ (1997) \\
& 0.0145 & 0.0020  & Thuan \& Izotov (1998) \\
& 0.017\phn & 0.0035  & Izotov \etal\ (1999) \\
& 0.0068 & 0.0013  & Peimbert \etal\ (2000) \\
& 0.0052 & 0.0009  & Olive \& Skillman (2001) \\
& 0.0170 & 0.0025  & Thuan \& Izotov (2002) \\
& 0.009\phn  & 0.002\phn   & Peimbert \etal\ (2002) \\
& 0.0095 & 0.0015  & Luridiana \etal\ (2003) \\
& 0.0125 & 0.0025  & Izotov \& Thuan (2004) \\
$^7$Li\tablenotemark{c} 
& 0.0146 & 0.0030  & Bonifacio \& Molaro (1997) \\
& 0.010\phn  & 0.004\phn   & Ryan \etal\ (2000) \\
& 0.015\phn  & 0.003\phn   & Vangioni-Flam \etal\ (2000) \\
& 0.0102 & 0.0038  & Suzuki \etal\ (2000) \\
& 0.016\phn  & 0.004\phn   & Bonifacio \etal\ (2002) \\
CBR: \LCDM\  
& 0.031\phn  & 0.0045  & Tegmark \& Zaldarriaga (2000) \\
& 0.022\phn  & 0.003\phn   & Netterfield \etal\ (2002) \\
& 0.024\phn  & 0.001\phn   & Spergel \etal\ (2003)\tablenotemark{d} \\
& 0.0224 & 0.0009  & Spergel \etal\ (2003)\tablenotemark{e} \\
CBR: no-CDM 
& 0.016\phn  & 0.006\phn & This work\tablenotemark{b} \\
\enddata
\tablenotetext{a}{$\ob \equiv \Ob h^2$.
  The calibration of Burles \etal\ (1999) has been used.}
\tablenotetext{b}{In these cases, $\pm \sigma$ gives the plausible range.}
\tablenotetext{c}{The higher \ob\ has been taken when double-valued.}
\tablenotetext{d}{Power law \LCDM\ fit.}
\tablenotetext{e}{\LCDM\ fit with a running spectral index.}
\end{deluxetable}

\clearpage

\begin{deluxetable}{lcccccccl}
\tablewidth{0pt}
\tablecaption{No CDM Models\label{nocdmmod}}
\tablehead{ \colhead{Model in} & \colhead{$\ob$} &
\colhead{$\Omega_{\Lambda}$} & \colhead{$\alpha$\tablenotemark{a}}
 & \colhead{$\fnu$} & \colhead{$n_t$} & \colhead{$\tau$}
 & \colhead{$A$\tablenotemark{b}} & \colhead{Ref.} }
\startdata
Fig.~3 (top) & 0.017 & 0.97\phn & 0.66 & 0.0\phn & 0.0\phn & 0.0\phn & 0.59 
 & M1999b \\
Fig.~3 (middle) & 0.019 & 1.006 & 1.0\phn & 0.0\phn & 0.0\phn & 0.0\phn & 0.61
 & M2000b \\
Fig.~3 (bottom) & 0.019 & 0.90\phn & 0.89 & 0.61 & 0.0\phn & 0.0\phn & 0.55
 &  \\
Fig.~4---8 & 0.019 & 0.918 & 1.0\phn & 0.65 & 0.04 & 0.17 & 1.01 & This work \\
\enddata
\tablenotetext{a}{Flat model scaled by $\LL \rightarrow \alpha \LL$
(McGaugh 2000b).}
\tablenotetext{b}{Amplitude normalization $C_{\LL} \rightarrow A C_{\LL}$.}
\end{deluxetable}

\clearpage 

\begin{deluxetable}{lccccl}
\tablewidth{0pt}
\tablecaption{Scaling Factors and $\chi_{\nu}^2$\label{fudgefact}}
\tablehead{ \colhead{Experiment} & \colhead{${\cal F}$\tablenotemark{a}} 
& ${\cal F}_{obs}$\tablenotemark{b} & \colhead{$\chi_{\nu}^2$} &
\colhead{$\chi_{\nu,\mathrm{mod}}^2$\tablenotemark{c}} &
\colhead{Modification}}
\startdata
BOOMERanG & 1.29 & 1.32 & 1.65 & 1.19 & Excludes $\LL = 700$ \\
MAXIMA        & \dots & \dots & 1.29 & \dots & \dots \\
DASI               & \dots & \dots & 1.95 & \dots & \dots \\
CBI                  & 0.76 & 0.94 & 1.25 & \dots & \dots \\
VSA	                & 0.95 & 0.92 & 1.76 & 0.93 & Excludes $\LL = 795$ \\
ACBAR           & 0.78 & \dots & 2.40 & 0.73 & Uses ${\cal F} = 0.78$ \\
ARCHEOPS    & \dots & \dots & 1.18 & \dots & \dots \\
WMAP             & \dots & \dots & 1.60 & 1.17 & Excludes $\LL = 40$ \\
ALL                  & \dots & \dots & 1.68 & 1.18 & All of the above \\
\enddata
\tablenotetext{a}{Scaling factor which  reconciles the calibrations
of each experiment: $C_{\LL} \rightarrow {\cal F} C_{\LL}$.}
\tablenotetext{b}{Observed correction to calibration postdating
the determination of ${\cal F}$.}
\tablenotetext{c}{$\chi_{\nu,\mathrm{mod}}^2$ shows the effect of
the noted modification.}
\end{deluxetable}

\clearpage

\begin{figure}
\plotone{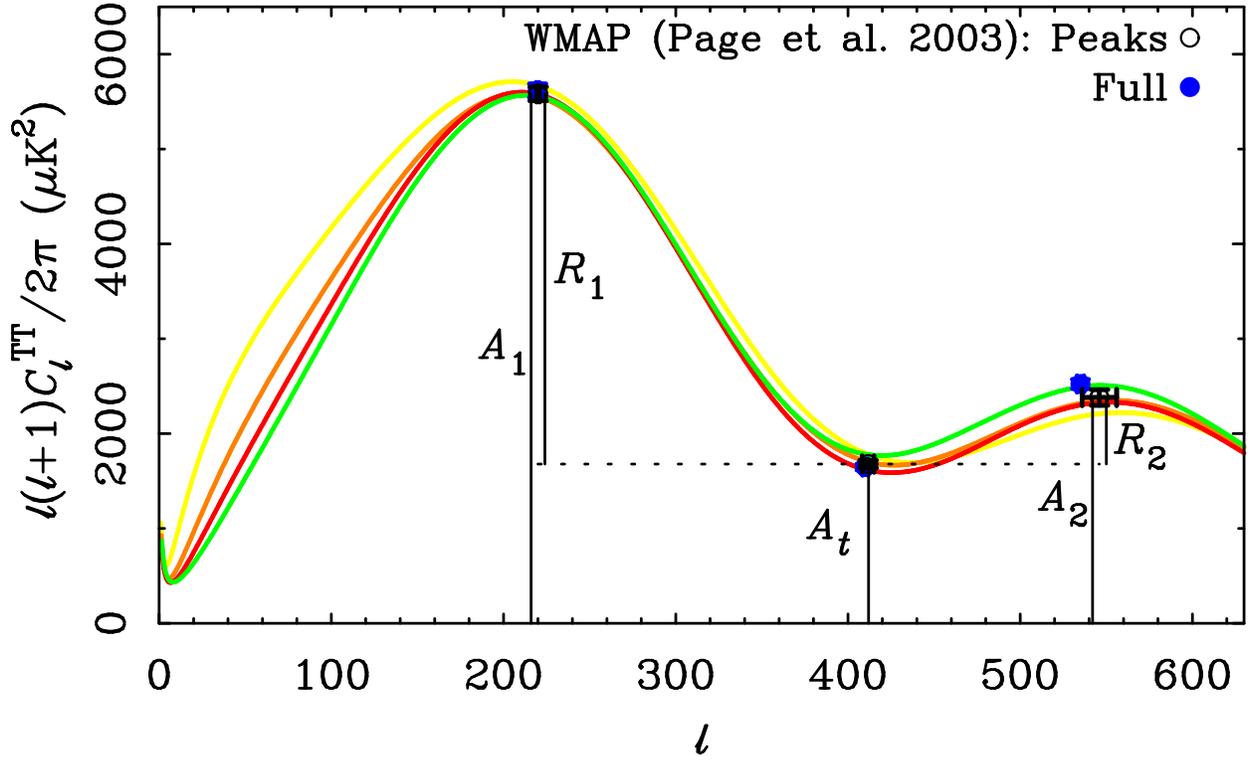}
\caption{The no-CDM models of McGaugh (1999b) are shown together with
the peak positions and amplitudes measured by WMAP as reported by
Page \etal\ (2003).  Also shown are the peak amplitudes used in defining
the ratios $\apk = A_1/A_2$ and $\rpk = R_1/R_2$.  No-CDM models
predict a very narrow range of these ratios for plausible (BBN) baryon 
densities, consistent with the WMAP observations.  The specific models
shown (from top to bottom at low \LL) are the $\Ob = 0.01$, 0.02, 0.03,
and $\On = \Ob = 0.02$ from McGaugh (1999b).
\label{f1}}
\end{figure}

\begin{figure}
\plotone{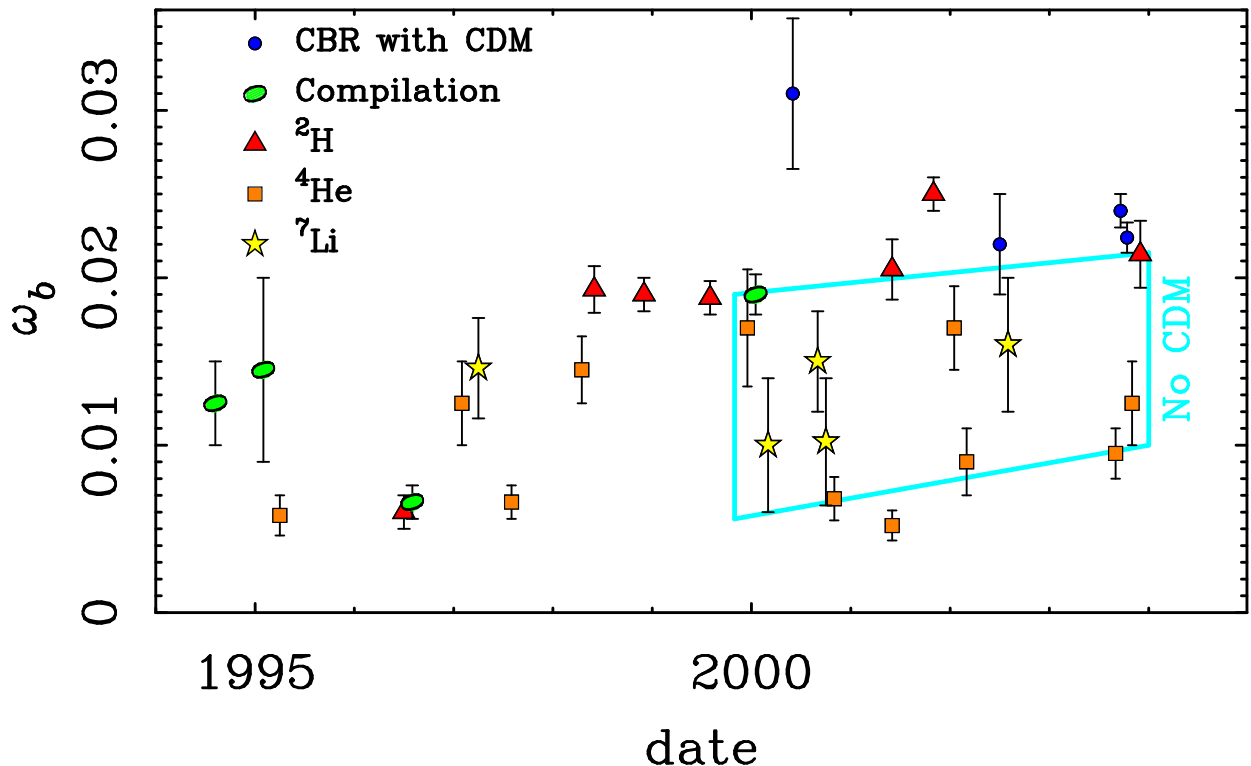}
\caption{Recent measurements of the baryon density $\ob = \Ob h^2$,
as given in Table~2.  The type of measurement is given by the various
symbols, including those works which are compilations of multiple isotopes.
BBN was a well developed field well before the start of this graph; earlier
work is represented by the compilation of Walker \etal\ (1991: leftmost point)
and Copi \etal\ (1995).  All error bars are $1 \sigma$ except as noted in
Table~2.  There is a dichotomy at $\wb = 0.020$.
The power law \LCDM\ fit from WMAP gives $\ob = 0.024 \pm 0.001$. 
Yet there were no BBN measurements which suggested $\ob > 0.020$ prior to the
first constraints from the CBR.  No-CDM models are consistent with a range
of lower \ob\ (box).  The left edge of the no-CDM box represents the range
of \ob\ considered by McGaugh (1999b), which was based on the plausible
range of Walker \etal\ (1991).  The right edge represents the range of
no-CDM models consistent with the WMAP first-to-second peak amplitude ratio. 
This range is consistent with independent estimates from all elements.
\label{BBNtime}}
\end{figure}

\begin{figure}  
\epsscale{0.75}
\plotone{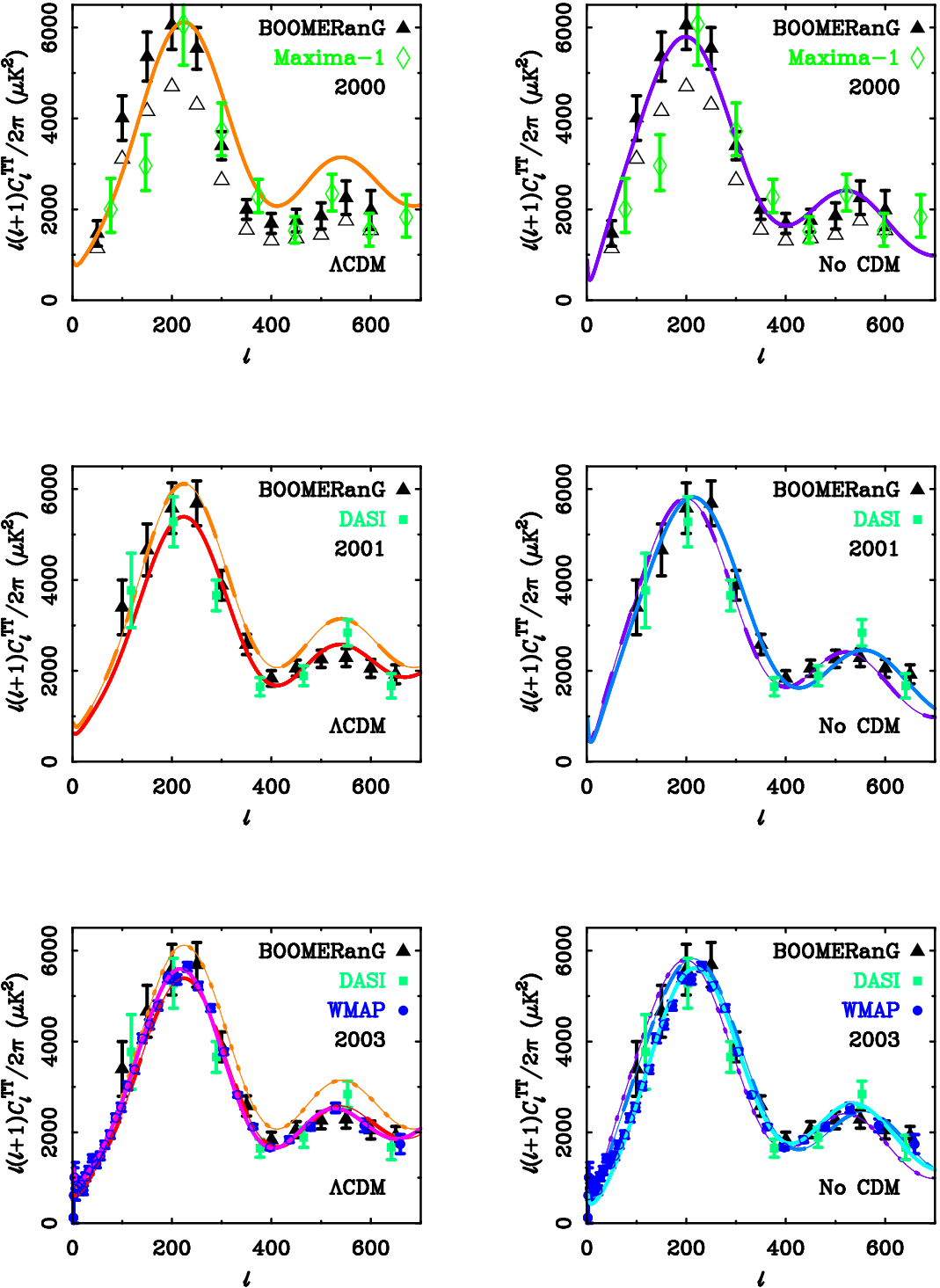}
\caption{The power spectrum of temperature fluctuations in the
microwave background, as observed by recent experiments.
Lines show model predictions and fits at the corresponding times,
with \LCDM\ models on the left and no-CDM models on the right.
The previous edition of each model is retained as dashed then
dotted lines in subsequent panels.
The no-CDM model provided a more accurate prediction of the
first-to-second peak amplitude ratio than did \LCDM, and it has
evolved considerably less in response to improving data.  All of
the no-CDM models shown in this figure existed before the
corresponding data (see text).
\label{f3}}
\end{figure}

\begin{figure}
\epsscale{1.0}
\plotone{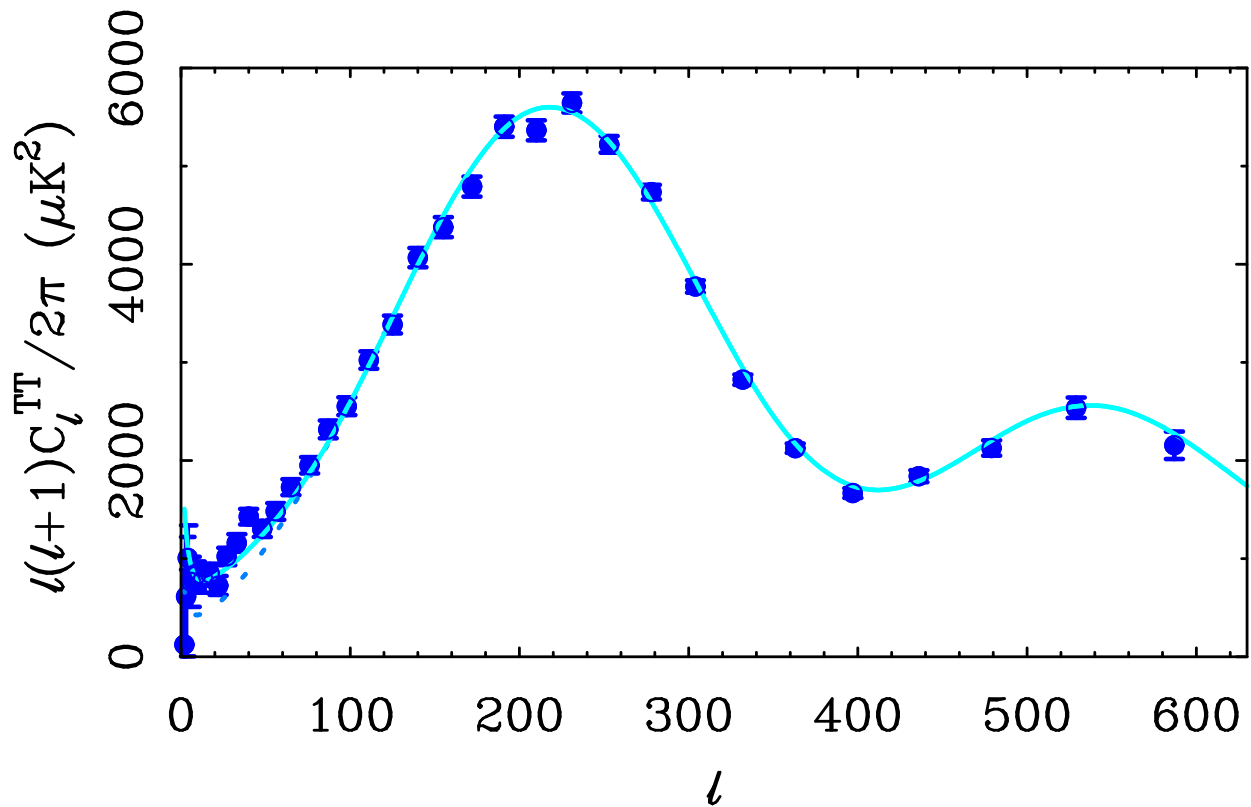}
\caption{A no-CDM model which matches the WMAP first year data.
The solid line is the model discussed in the text.
The dashed line is the closest pre-existing no-CDM model (Table~3). 
The two are indistinguishable for $\LL > 100$.
\label{f4}}
\end{figure}

\begin{figure}
\plotone{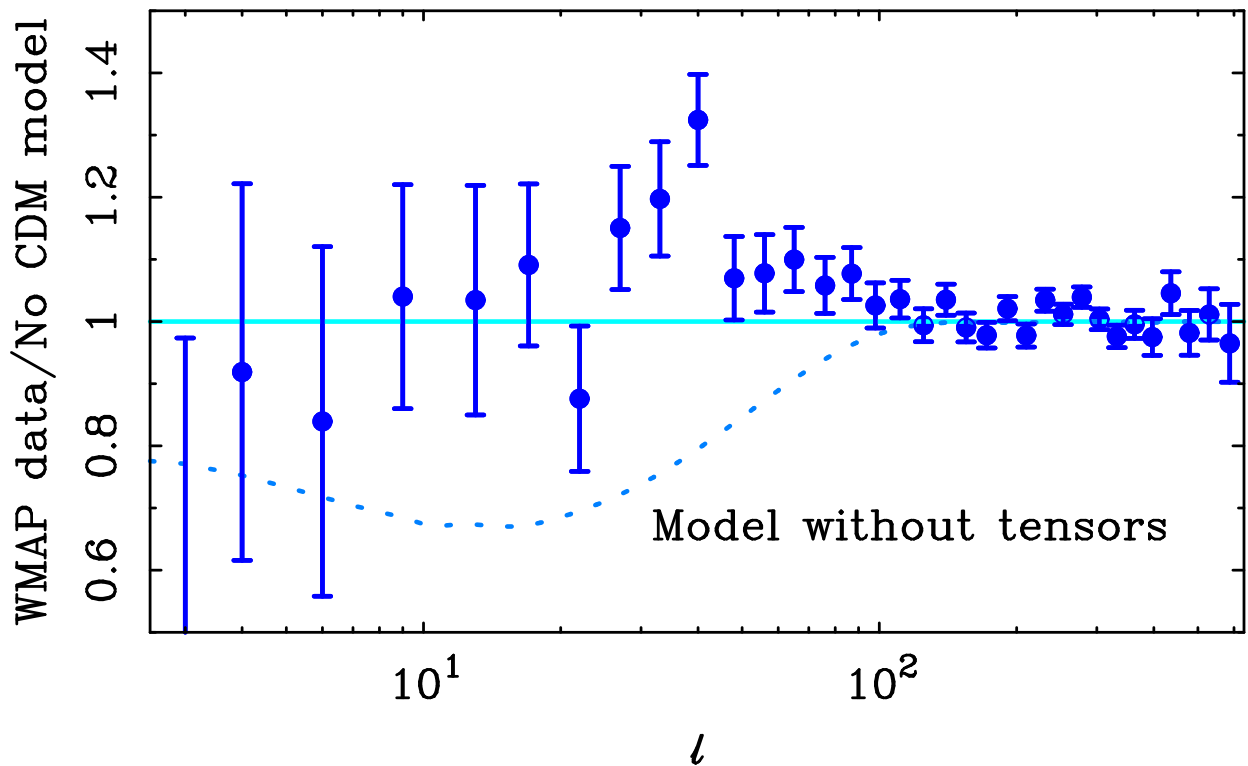}
\caption{The ratio of WMAP data to the best fitting no-CDM model,
with a logarithmic abscissa to emphasize small $\ell$.  
The scalar-only no-CDM model (dotted line)
fits well for $\ell > 100$, but systematically under-predicts the
power at small $\ell$.  The excess power suggestive of an
amplified ISW effect.  This is one predicted signature of MOND-like
physics beyond the simple no-CDM model.  
\label{f5}}
\end{figure}

\begin{figure}
\epsscale{1.0}
\plotone{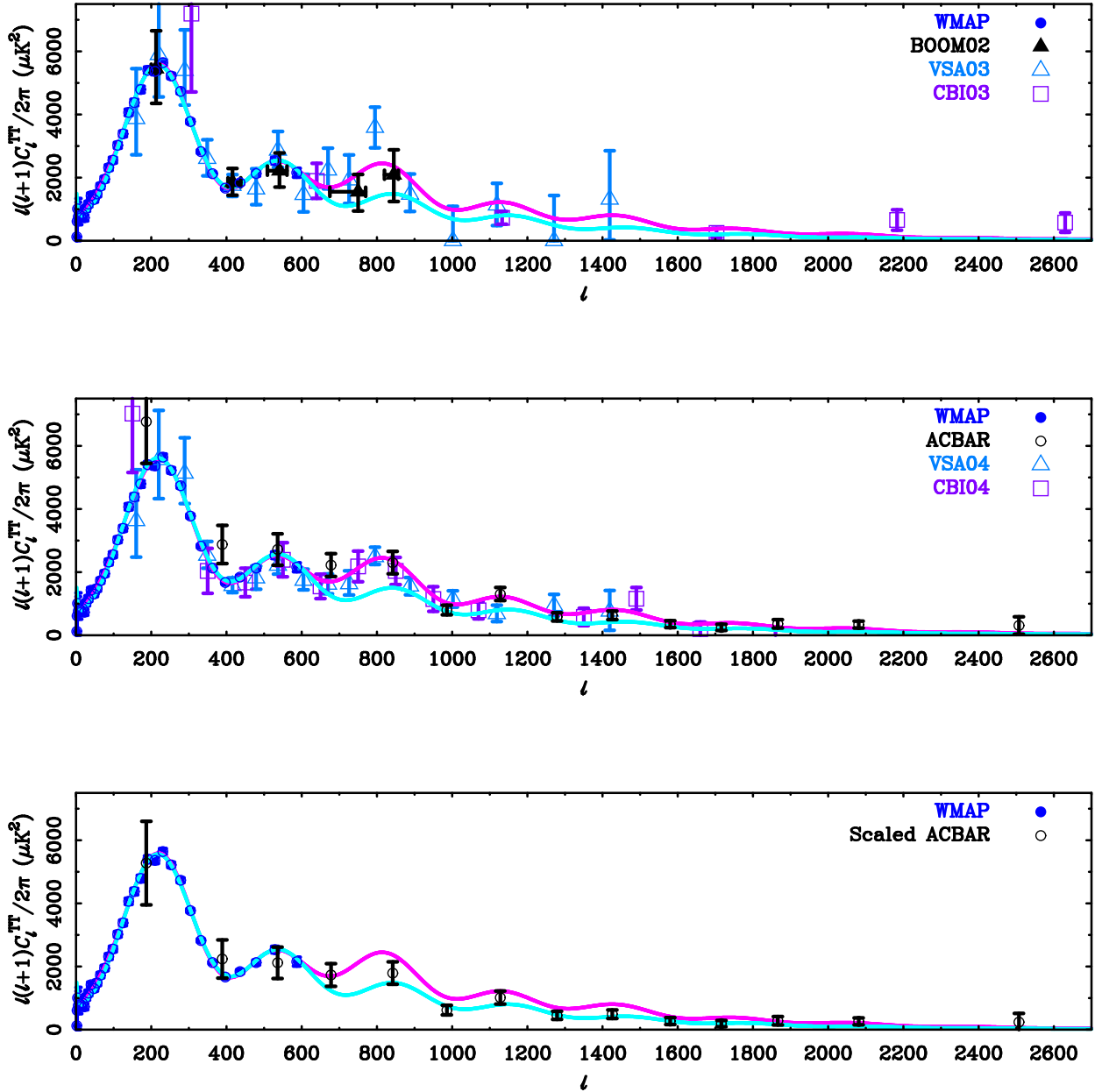}
\caption{The CBR power spectrum extending to large \LL.  
The top panel shows data from CBI and VSA, as reported in 2003,
together with the WMAP data and the BOOMERanG peak locations.
The middle panel shows the ACBAR data
and the updated (2004) versions of the VSA and CBI data.
The lower panel shows the ACBAR data scaled to match the
WMAP data in the range of \LL\ where the two overlap.  
Note that the ACBAR data closely follow the predictions of the model
lines, though which one depends on the scaling.  The lines are the
power law WMAP \LCDM\ model and the no-CDM model. 
These are indistinguishable for $\LL < 600$. 
The models diverge for $\LL > 600$, with \LCDM\ predicting the larger
third peak and no-CDM the smaller.
\label{f4}}
\end{figure}

\begin{figure}
\epsscale{1.0}
\plotone{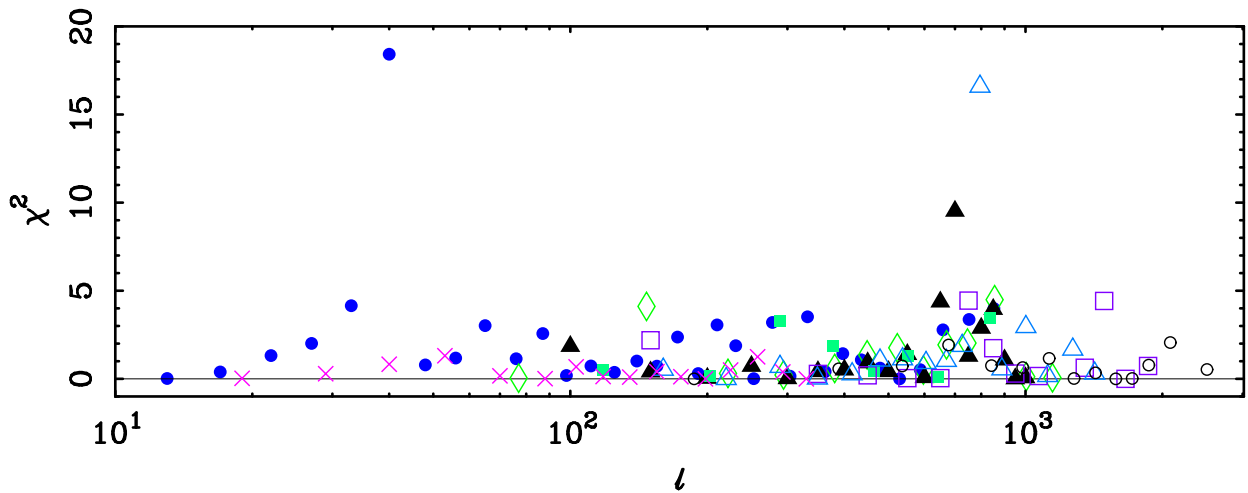}
\caption{$\chi^2$ of individual data points for each of the experiments
listed in Table~4.  Symbols are the same as in previous figures.  
ACBAR is plotted scaled with ${\cal F} = 0.78$.
The $\chi^2$ budget is dominated by the WMAP point at $\LL = 40$
and the VSA point at $\LL = 795$.  There is a clear indication of
power in excess of the no-CDM prediction in the vicinity of the
third peak, but only the single point form VSA is highly significant.
\label{chisq}}
\end{figure}

\begin{figure}
\plotone{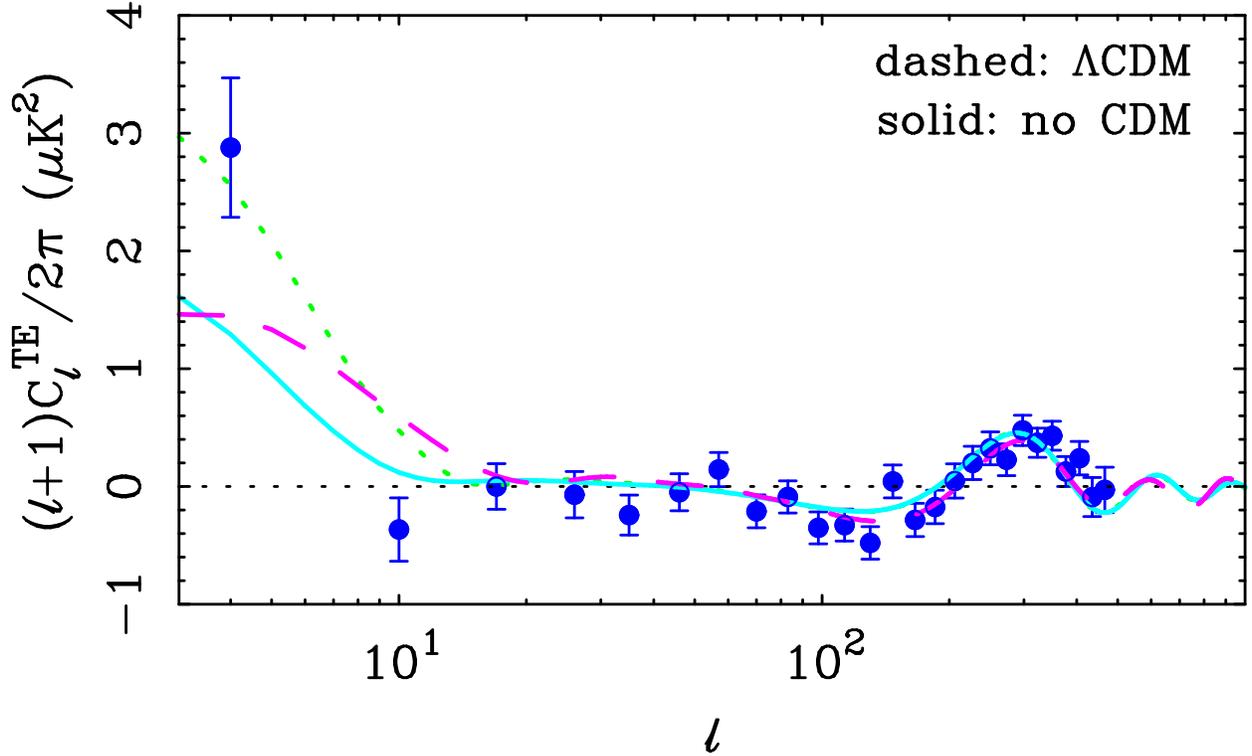}
\caption{The TE polarization signal in the WMAP data (points).
Also shown is the \LCDM\ model form the WMAP team (dashed line) and two
no-CDM models.  The solid line is the same model as in previous plots,
with optical depth fixed to $\tau = 0.17$.  The dotted line is a similar model
with $\tau = 0.30$.  This is at the upper edge of the confidence interval
quoted by Kogut \etal\ (2003), but is not obviously inconsistent with
the data in this plot.  An early epoch of structure formation, leading
to relatively high optical depth, is natural in MOND and was one of
the {\it a priori\/} predictions of McGaugh (1999b).
\label{f5}}
\end{figure}

\end{document}